\documentclass[final,authoryear,5p,times,twocolumn]{elsarticle}
\pdfoutput=1

\usepackage{graphicx}

\usepackage{amssymb}

\usepackage[pdftex,pdfpagemode={UseOutlines},bookmarks,bookmarksopen,colorlinks,linkcolor={blue},citecolor={green},urlcolor={red}]{hyperref}
\usepackage{hypernat}

\journal{Astronomy \& Computing}

\usepackage{upquote}

\usepackage{upgreek}

\newcommand{\locator}{\emph{locator}}
\newcommand{\locators}{\emph{locators}}
\newcommand{\new}{HDSv5}
\newcommand{\old}{HDSv4}

\begin{document}

\begin{frontmatter}



\title{Reimplementing the Hierarchical Data System using HDF5}


\author[cornell]{Tim Jenness\corref{cor1}}
\ead{tjenness@cornell.edu}

\cortext[cor1]{Corresponding author}

\address[cornell]{Department of Astronomy, Cornell University, Ithaca,
  NY 14853, USA}

\begin{abstract}
The Starlink Hierarchical Data System has been a very successful niche
astronomy file format and library for over 30 years. Development of the library
was frozen ten years ago when funding for Starlink was stopped and almost no-one
remains who understands the implementation details. To ensure the
long-term sustainability of the Starlink application software and to
make the extensible N-Dimensional Data Format accessible to a
broader range of users, we propose to re-implement the HDS library
application interface as a layer on top of the Hierarchical Data
Format version 5. We present an overview of the new implementation of
version 5 of the HDS file format and describe differences between the
expectations of the HDS and HDF5 library interfaces. We finish by
comparing the old and new HDS implementations by looking at a
comparison of file sizes and by comparing performance benchmarks.

\end{abstract}


\end{frontmatter}


\newcommand{\mnras}{MNRAS}
\newcommand{\aap}{A\&A}
\newcommand{\aaps}{A\&AS}
\newcommand{\pasp}{PASP}
\newcommand{\apj}{ApJ}
\newcommand{\apjs}{ApJS}
\newcommand{\qjras}{QJRAS}
\newcommand{\an}{Astron.\ Nach.}
\newcommand{\ijimw}{Int.\ J.\ Infrared \& Millimeter Waves}
\newcommand{\procspie}{Proc.\ SPIE}
\newcommand{\aspconf}{ASP Conf. Ser.}


\newcommand{\KAPPA}{\textsc{kappa}}
\newcommand{\gaia}{\textsc{gaia}}
\newcommand{\figaro}{\textsc{figaro}}
\newcommand{\ccdpack}{\textsc{ccdpack}}
\newcommand{\smurf}{\textsc{smurf}}
\newcommand{\surf}{\textsc{surf}}
\newcommand{\asterix}{\textsc{asterix}}
\newcommand{\specdre}{\textsc{specdre}}
\newcommand{\iras}{\textsc{iras90}}
\newcommand{\treeview}{\textsc{treeview}}
\newcommand{\splat}{\textsc{splat}}
\newcommand{\catpac}{\textsc{catpac}}
\newcommand{\CFITSIO}{\textsc{cfitsio}}
\newcommand{\fitstondf}{\textsc{fits}{\footnotesize{2}}\textsc{ndf}}

\newcommand{\ascl}[1]{\href{http://www.ascl.net/#1}{ascl:#1}}

\section{Introduction}
\label{sec:intro}

The Hierarchical Data System (HDS) was created by the Starlink Project
in the United Kingdom in the early 1980s
\citep{1982QJRAS..23..485D,1991STARB...8....2L}. The requirements were
to have a file format that was optimized for data processing
applications: allowing for efficient access of data arrays through
memory mapping, grouping of related data structures in a hierarchy to
make it easy to move data \emph{en masse} from one location to
another, and easy modification of data and structures. At the time
FITS \citep{1981A&AS...44..363W} was mainly thought of as a transport
format distributed on tape \citep{1980SPIE..264..298G}, and the NCSA
Hierarchical Data Format would not be developed until the end of the
decade \citep{HDF1,Folk2010}. It was therefore decided to develop a
new file format from scratch. Initially called the Starlink Data
System before being rebranded as HDS, the first version of the library
was written in BLISS-32 before being ported to C on VAX/VMS in the
late 1980s \citep{SSN27a}.

HDS succeeded in its goal of forming the basis of the Starlink data
reduction software packages (\ascl{1110.012}) and was and is being
used at UK observatories for data acquisition and for data reduction
pipelines \citep{2011tfa..confE..42J,2015A&C.....9...40J,2014ASPC..485..143B}. Its presence was
pervasive within the Starlink software stack, being used for parameter
storage in the ADAM system \citep{1992ASPC...25..126A} and in a
graphics database system \citep{SUN48} in addition to storing
astronomy data and forming the basis of the Starlink
\emph{N}-Dimensional Data Format library \citep[NDF;][\ascl{1411.023}]{2015Jenness}. There was
very little take up of the format outside the UK community \citep[but
see e.g.,][]{1993STARB..12...10M} and as FITS came to be used as a data
processing format as well as a transport and archive format, HDS has become a niche product.

HDF5 is a popular file format in other scientific disciplines and is
used in fields such as Earth science \citep[e.g.,][]{2005Yang}, biology
\citep[e.g.,][]{Dougherty:2009:UBI:1562764.1562781}, nuclear physics
\citep[e.g.,][]{1742-6596-425-6-062008}, and molecular simulations
\citep[e.g.,][]{deBuyl20141546}. The astronomy community is currently
discussing the wider issues of file formats beyond FITS
\citep{B1_adassxxiv,2015Thomas} and HDF5 is being adopted
\citep[e.g.,][]{2012ASPC..461..283A} or investigated
\citep[e.g.,][]{P3-1_adassxxiv,O4-4_adassxxiv} in a number of astronomy
projects.

Given this context it is therefore worth investigating whether there
should be a new version of HDS that is based on HDF5. In this paper we
compare the HDS and HDF5 data models, discuss the motivations for such
a change, and describe an implementation.

\section{Motivation}

There are a number of key motivators for migrating from the current HDS
format to a more widely-recognized format:
\begin{enumerate}
\item Opaque implementation details of the library and format with no
  resident expert or associated documentation.
\item Lack of support for 64-bit dimensions sizing.
\item HDS has no provision for transparent data compression.
\item HDS has no native support for tables.
\item Sociological impediment to adopting a niche format in the wider
  astronomy community.
\end{enumerate}
We will discuss each of these in turn.

\subsection{Opaque implementation}

The Starlink Project was closed in 2005 after 25 years of
operation. The Starlink Software Collection continues to be developed
as an open-source project sponsored by the Joint Astronomy Centre to
continue support for their data reduction pipelines. Unfortunately
none of the remaining developers understand the implementation details and there is
no incentive for anyone to learn given other development priorities. The most recent description of the
internals of the HDS file format is in \citet{SSN27a} which documents
version 2 of the format. Version 3 (from the port to Unix in 1991) and
version 4 (in 2005 to support files larger than 2GB) remain
undocumented, outside of extensive comments in the code itself,
with an assumption that the design closely matches the
original layout. For a data format that is used as an archive format
\citep[see e.g.,][]{2015Economou} this lack of documentation and understanding
represents a risk for long-term access to the data.

\subsection{64-bit dimension sizes}

The HDS library does not currently support 64-bit dimension
sizes. This was partially implemented for version 4 of the format but
was never completed. It is not clear how much work it would be to finish
this work or whether anyone remaining can do so. As data rates
increase it is clear that the next generation of heterodyne arrays
\citep[e.g.,][]{2014SPIE.9152E..2WJ} will generate data cubes that exceed
the capacity of a 32-bit integer counter and these files will render
the Starlink software and associated data pipelines unusable without
HDS being upgraded. Adding 64-bit support to HDS is a necessary but
not sufficient step towards the applications supporting larger
datasets, and most libraries and applications in Starlink will
need to be updated to support 64-bit counters. HDS is the
fundamental building block that has to be converted first, providing
new APIs to allow both 32-bit and 64-bit application code to co-exist.

\subsection{Data Compression}

HDF5 supports transparent data compression of datasets using a number
of algorithms in addition to supporting a pluggable architecture. HDS
does not support any data compression and relies on the facilities of
the NDF library to provide these. NDF can support gzipped data files,
requiring a temporary file, as well as FITS-style
\texttt{BSCALE}/\texttt{BZERO} and a delta compression scheme
natively. Leveraging the HDF5 compression algorithms would be a very
easy way to improve the data compression performance in NDF.

\subsection{Native Tables}

Unlike FITS \citep{1988A&AS...73..365H} or HDF5, HDS does not have a
native table data structure. Tables must be implemented as a
collection of independent columns and this can be extremely
inefficient for row access. Switching data formats would make it
possible to add a native table access API to HDS.

\subsection{Sociology}

One of the impediments to adoption of the NDF data model is that the
reference library implementation uses HDS as the underlying file
format. Whilst the NDF data model itself does not require HDS and
could be implemented by anyone with sufficient effort available, to
adopt NDF currently requires that users adopt HDS.
Adopting HDF5 would considerably lower the barrier to entry for people
more comfortable in the HDF5 world or who are considering switching
from another format, and make
data files accessible to tools such as \texttt{HDFview} and
\texttt{h5py}. Of course, none of these tools will understand the NDF
data model that defines the hierarchical grouping but it makes it
easier for other tools to adopt some of the same conventions.
Similarly, the Starlink file format conversion tools \citep{SUN55} would be able
to import formats such as FITS to HDF5 and this infrastructure may
prove to be useful for people who are themselves switching to HDF5.

\section{Features of HDS}

How HDS is used depends on a data model and the data access
model. Both of these are important when considering a change in
implementation.

\subsection{Data Model}

\begin{table*}[!ht]
\caption{HDS basic data types. The unsigned types did not correspond
  to standard Fortran~77 data types and were included for
  compatibility with astronomy instrumentation. HDS supports both VAX
  and IEEE floating-point formats. The API code indicates the letter appended
  to function names to indicate the type they support. This convention is
  used for the \textsc{generic} templating system \citep{SUN7}. For
  compatibility with HDS the \_LOGICAL type is a 32-bit bitfield type
  in memory but stored in an HDF5 file using 8 bits. HDS strings are
  always stored in Fortran space-padded form and that convention is
  adopted in the HDF5 HDS implementation.}
\label{tab:hdstypes}
\begin{center}
\begin{tabular}{llll}
\hline
Name of type & API Code & Data type & HDF5 Data type  \\ \hline
\texttt{\_BYTE} & b & Signed 8-bit integer & H5T\_NATIVE\_INT8 \\
\texttt{\_UBYTE} & ub & Unsigned 8-bit integer & H5T\_NATIVE\_UINT8\\
\texttt{\_WORD} & w & Signed 16-bit integer &   H5T\_NATIVE\_INT16\\
\texttt{\_UWORD} & uw & Unsigned 16-bit integer &  H5T\_NATIVE\_UINT16 \\
\texttt{\_INTEGER} & i & Signed 32-bit integer &  H5T\_NATIVE\_INT32 \\
\texttt{\_INT64} & k &Signed 64-bit integer & H5T\_NATIVE\_INT64 \\
\texttt{\_LOGICAL} & l & Boolean & H5T\_NATIVE\_B8 \\
\texttt{\_REAL} & r & 32-bit float & H5T\_NATIVE\_FLOAT \\
\texttt{\_DOUBLE} & d & 64-bit float & H5T\_NATIVE\_DOUBLE \\
\texttt{\_CHAR[$*$n]} & c & String of 8-bit characters  & H5T\_STRING\\
\hline
\end{tabular}
\end{center}
\end{table*}

HDS is a hierarchical file format where named structures can contain
other named structures or named primitive data arrays. It is
self-describing in the sense that each layer in the hierarchy can be
queried to obtain the number of members below (for structures), their
names and their types.

The primitive objects support numerical and string types with up to 7
dimensions. The supported data types are shown in
Table\,\ref{tab:hdstypes} and the choice and names reflect the origins
of the format as a library designed to be used from Fortran.

All HDS components are typed and this includes structures. Structure
typing is important in HDS as it can be used by an application to
decide whether a structure can be understood or not. In object-oriented
nomenclature the type can be thought of as the object class, whereas the
individual named structure is an object instance.

HDS supports the concept of arrays of structures to allow a collection
of identical structures to be grouped together. This is used
extensively in the lower-levels of the Starlink software stack, for example to
support history recording (an array of structures of type
\texttt{HISTORY} which is extended each time a new history record is
created), or picture definitions in the graphics database.

\subsection{Data Access Model}

To obtain access to an object within an HDS file the caller must
obtain a \locator; an opaque C \texttt{struct} containing information
about the object needed by the HDS library. These \locators\ mediate
all access to the HDS data file.

\paragraph{Component copying}

As a consequence of the hierarchical design, it is possible to copy or
move arbitrary parts of the tree to other locations within a file or
locations in different files.

\paragraph{Primary and secondary locators}

The library has a concept of primary and secondary \locators\ such that
when all primary \locators\ associated with a file are freed (or
\emph{annulled} in HDS parlance) all resources associated with that
file are also closed and all secondary locators become inactive.

\paragraph{Locator Groups}

It is possible to assign locators to a named group. Any child locators
are also members of the group. When the group is no longer required it
can be flushed with a single command, freeing all the locators that
are in the group. This simplifies the management of large numbers
of related locators and allows the resources to be freed at one place
in the code without having to store them all in user code.

\paragraph{Slicing}

Arrays of structures can not be accessed directly but must instead be
accessed by requesting a specific \emph{cell}. Primitive data arrays
can also be accessed by individual cells but it is more common to
access data arrays by specifying slices. A slice can be requested by
specifying upper and lower bounds of each dimension. The Fortran
heritage requires that these bounds are indexed starting with a lower
bound of 1 rather than 0, and all data arrays are specified in Fortran
order; even from the C interface. In some cases the dimensionality is
unimportant and the library allows a \locator\ to be vectorized such
that subsequent interrogations of the locator will indicate that the
object is 1-dimensional regardless of the underlying shape. This can
be very useful for such activities as examining every element in turn,
or picking the first few elements. Vectorizing works for structures
and primitives and does not affect the file itself.

\paragraph{Automatic type conversion}

For primitive arrays, the data to be stored or the data to be
retrieved do not have to be the same type as the format of the data
stored on disk. Floating point data will be converted to integer and
vice versa. Also, string and logical/boolean types will be converted
to numbers and numbers can be retrieved as strings or
logicals. Endianness and floating point representation is also handled
transparently, and the native form is used when a file is created.

\paragraph{Memory Mapping}

One of the initial requirements for HDS was efficient access to data
arrays. This was done using direct mapping of the relevant part of the
file into memory\footnote{Using \texttt{mmap()} on POSIX systems} and
was implemented for read and write operations. The emory mapping facility
can be enabled or disabled by use of an environment variable and an
in-memory solution is used on systems that do not support memory mapping.

\section{Requirements for an Updated Format}

The Starlink software collection consists of more than 2.3 million
lines of Fortran, C and C++ and a large fraction of that code depends
on the HDS library and the HDS API. This includes fundamental
infrastructure such as ADAM that is used by all applications. It is
therefore imperative that the API for HDS remains the same even with
the implementation changing underneath. Any new version of HDS should
meet the following requirements:

\begin{enumerate}
\item The API should not change.
\item It should be possible to use both old and new format files in
  the same application.
\item The application should behave in the same way with new files as
  it does with old files.
\item The application source code should not need to be modified in
  any way to use the new library.
\item The new format should not impact performance of the application
  in a negative way or require more computer resources.
\end{enumerate}

These are similar to the requirements described when NetCDF version 4
was implemented on top of HDF5 \citep{2004Rew}.

\section{HDS Version 5}

Given the broad adoption of HDF5 in the scientific community and the
close similarity in key parts of the data model between it and HDS, it
was decided to write a prototype implementation of the HDS API in
terms of HDF5. This would provide information on the feasibility of
the approach and also highlight the areas where the data models or
access models diverge. The previous version of HDS\footnote{Somewhat
  confusingly the library implementing version 4 of the file format is
  itself version 5} was version 4 so it
was decided this version would be version 5.
In the rest of this paper we use the shorthand
\new\ to refer to the new library implementation and format, and \old\
to refer to the current version of the HDS format and library.

\subsection{Library Architecture}

\begin{figure}[t]
\begin{center}
\includegraphics[width=0.7\columnwidth]{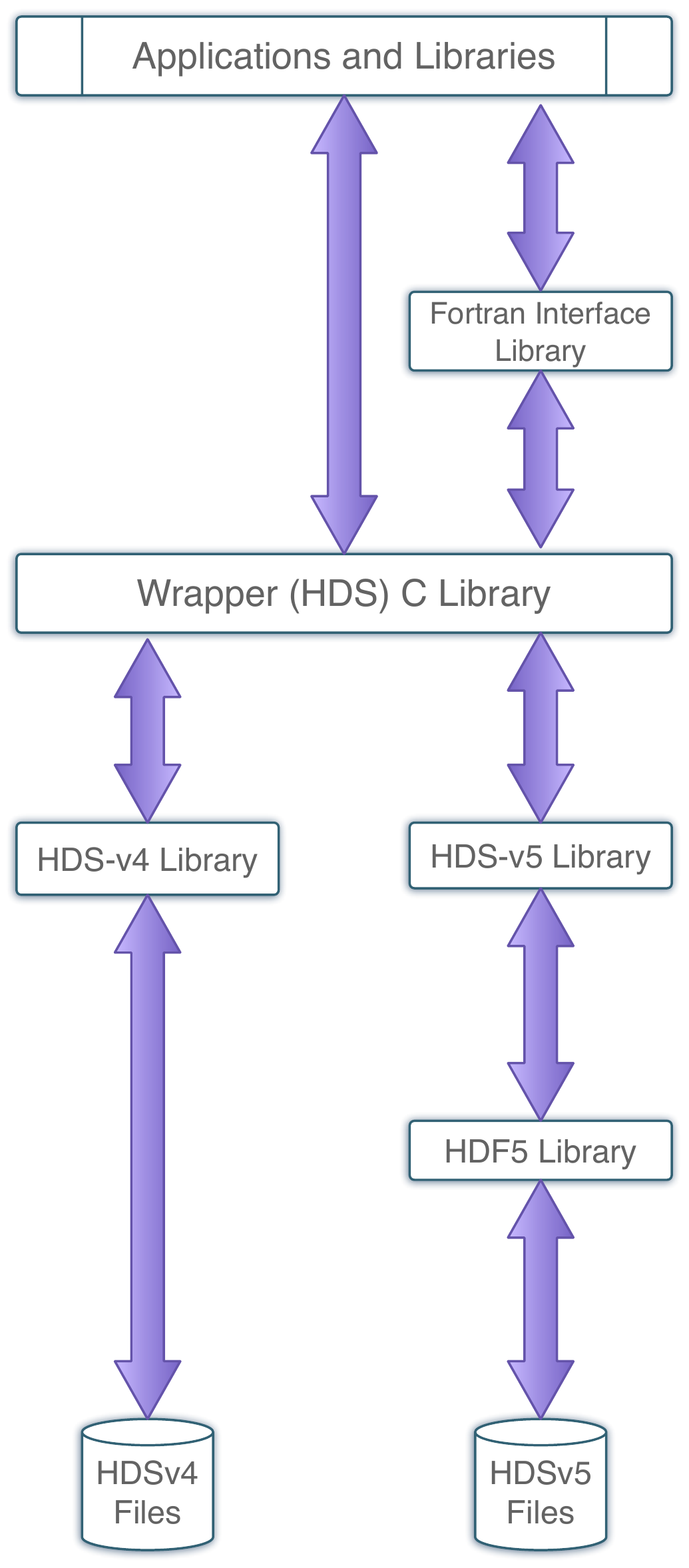}
\end{center}
\caption{Architecture of the HDF5-based implementation of HDS. A
  wrapper library with the public HDS API forwards calls to the
  correct version of the library. The Fortran interface is a separate
  library as it also contains Fortran code that would require a
  Fortran runtime library.}
\label{fig:arch}
\end{figure}

In order to support both new and old file formats it was necessary for
the new library to have access to a complete copy of the existing
library. The HDF5-based library and the \old\ library are both
standalone libraries that are linked in to a wrapper library that
implements the public interface (see Fig.~\ref{fig:arch}). The versioned libraries can be
configured to provide the public API but when used as part of the
unified wrapper they are built with names that include the version
number to avoid symbol clashes.

The wrapper library is responsible for forwarding the calls to the
correct underlying library. There are four major API styles that must
be handled: functions that open files and return locators, functions
that create files, functions that copy from one locator to another,
and functions that work with a single locator.

When a request is made to open a new file, it is first sent to the
\old\ library to see if it opens. If that fails due to the format
being invalid, the \new\ library is used to open the file. When
migration to the new format is substantially complete the wrapper will
be modified to default to using the \new\ library first. One caveat is
that the library must ensure that \new\ files are written to disk
immediately on creation\footnote{Calling \texttt{H5Fflush} in
  \texttt{hdsNew}} such that the HDF5 superblock signature is written.
Without this step \texttt{H5Fis\_hdf5} will not correctly determine
that a newly created file is an HDF5 file if it has not yet been
closed and some Starlink applications and libraries rely on the
ability to create an HDS file and then open it in another part of the
code without having annulled all previous locators beforehand.

When files are to be created the choice of format is controlled by a
tuning parameter. Tuning parameters in HDS can be set programmatically
or by reading the environment. By default, files are still created in
\old\ format using the principle of least surprise. The ability to
control this behavior from an environment variable simplifies testing
and benchmarking.

When copying one locator to another locator of a different type,
tree-walking code had to be written using the HDS public API. The code
recursively walks through structures copying primitives and other
structures as required.

The bulk of the API takes a single locator as input and does something
with it that may or may not result in a new locator being created. We
have taken a slightly different approach to that described in
\citet{2004Rew}. In that paper they registered function table lookups
with the newly created data objects, allowing efficient forwarding to
the particular library. We decided to take a simpler approach whereby
the locator structures in \old\ and \new\ were adjusted so that they
both included a version integer as the first member. The wrapper code
then simply checks for the version number in the structure and calls the
relevant routine. This approach does simplify the addition of
debugging messages and error reporting from each routine at the
expense of some calling efficiency.

The wrapper code responsible for this forwarding is generated from the
public HDS header file using a simple Python program. This allows the
forwarding scheme to be changed relatively easily.

\subsection{The Locator Interface}

As mentioned previously, a \locator\ is an opaque C \texttt{struct}
containing information about a particular object in the HDS file. The
size of of these structures differs between the two
implementations and this required that a change be made to the Fortran
interface of the HDS library. For historical reasons an HDS \locator\
is stored in a Fortran character array with a length specified by the
HDS constant \texttt{DAT\_\_SZLOC} (currently 16 characters). In
\old\ the C structure is really a proxy for an internal data
structure and the Fortran interface copied the contents of the
structure to and from the Fortran string. In \new, the C
structure is significantly larger and it was unreasonable to increase
the Fortran \locator\ size. To support both implementations the
Fortran interface was changed such that the structure address was
stored in the Fortran string buffer and the size of the string was
kept at 16 characters. This allowed the new library to be installed
without requiring that any applications be relinked.

\subsection{Error handling}

HDS uses the Starlink Error Message Service \citep[EMS;][]{SSN4} for
error handling. EMS uses the concept of inherited status where each
function takes a status argument and usually returns immediately if
status is non-zero (when resources are to be freed it is usual for the
freeing routine to try to execute regardless). If an error condition
is to be reported by a function it sets the status to an appropriate
value and attaches an error message to the message stack. As the call
stack unwinds the error could either be annulled (the calling function
may wish to react to the error by trying an alternative) or be
augmented with more information.

HDF5 uses a similar error message stack and status code concept
internally but uses a function return value to indicate to the
external user that a problem has occurred. If the return value is
negative the call failed. The error messages and specific status code
must then be retrieved separately. In \new\ each call to HDF5 is
wrapped by a C macro that intercepts the status return and if
necessary queries the HDF5 error message stack and places each of the
messages on to the EMS message stack.

In some cases the HDF5 status code is translated to an equivalent HDS
error code but in many cases the HDF5 codes are not specific enough
and in that case a generic \emph{error from HDF5} code is used.

\subsection{Data Model}

In HDF5, structures are known as \emph{groups} and primitives are
known as \emph{datasets}. Table\,\ref{tab:hdstypes} shows the mapping
of HDS data types to the HDF5 equivalents and the type system is
significantly more advanced in HDF5. It was decided that boolean types
should be represented in the files as 8-bit bitfields rather than the
32-bit integer type that is used (part of the Fortran legacy). The
in-memory datatype is a 32-bit integer for consistency with the public
API but the smaller type is used on disk. A bitfield type is used as
this allows the HDS type query to be able to distinguish the
\texttt{\_BYTE} type from \texttt{\_LOGICAL} type without requiring
the use of HDF5 attributes. Strings in HDS are space-padded fixed size
following the Fortran style and this is how strings are stored in
\new. Datasets are stored in HDF5 files in C dimension order with the
dimensions being reversed when viewed from HDS. This is the same
approach taken by the HDF5 Fortran interface with the variation that
the HDS C view of an array must agree with Fortran.

HDF5 has no concept of arrays of structures so this facility is
implemented entirely by the \new\ library. The containing group is
created and within it are placed the number of groups corresponding to
the array size. Each of these groups is given a name that contains a
root string chosen to deliberately be longer than the maximum allowed
length of an HDS component, appended with the coordinates of the
structure in the array. For a 2-dimensional array of structures the
name of the group could be \texttt{ARRAY\_OF\_STRUCTURES\_CELL(2,3)}
for the group at coordinate (2,3). This naming scheme simplifies
access to an individual structure (just provide the coordinates) and
also simplifies reporting of the full path using HDS nomenclature: to
convert the HDF5 path of the structure
\texttt{ROOT/HISTORY/ARRAY\_OF\_STRUCTURES\_CELL(3)} to the HDS path,
just requires the removal of the fixed cell prefix to convert it to
\texttt{ROOT.HISTORY(3)}\footnote{HDS uses dot separators rather than
  directory separators when specifying a path within a data file. This
  will be familiar to VMS users.}. The
long structure name is hidden by the HDS library and only visible when
the file is accessed using HDF5 tools. When an array of structures has
been created the dimensionality is stored in an attribute named
\texttt{HDS\_STRUCTURE\_DIMS}. In the future we will consider
implementing structure arrays using the HDF5 feature allowing
references to arbitrary HDF5 objects to be stored in a dataset, this
would have the advantage of reducing the structure complexity and
would simplify cell access.

Finally, the data type of a structure is not a fundamental part of
HDF5 so this information is stored in an attribute with name
\texttt{CLASS} following the convention used in other HDF5 data models
such as the Image and Palette classes.

Fig.~\ref{fig:compare} shows a comparison of the HDS and HDF5 view of
the same data file.  These traces show that the mapping from HDS
structure/primitive to HDF5 group/dataset is being followed with three
attributes added to provide the metadata required by HDS.

\begin{figure}[!ht]
\small
(a)
\begin{verbatim}
IMAGE  <NDF>
  DATA_ARRAY  <ARRAY>     {structure}
    DATA(2)     <_INTEGER>   1,2
\end{verbatim}
(b)
\begin{verbatim}
GROUP "/" {
   ATTRIBUTE "CLASS" {
      DATATYPE  H5T_STRING {}
      DATASPACE  SCALAR
      DATA {
      (0): "NDF"
      }
   }
   ATTRIBUTE "HDS_ROOT_NAME" {
      DATATYPE  H5T_STRING {}
      DATASPACE  SCALAR
      DATA {
      (0): "IMAGE"
      }
   }
   GROUP "DATA_ARRAY" {
      ATTRIBUTE "CLASS" {
         DATATYPE  H5T_STRING {}
         DATASPACE  SCALAR
         DATA {
         (0): "ARRAY"
         }
      }
      DATASET "DATA" {
         DATATYPE  H5T_STD_I32LE
         DATASPACE  SIMPLE { ( 2 ) / ( 2 ) }
         DATA {
         (0): 1, 2
         }
      }
   }
}
\end{verbatim}
\caption{Comparison of the HDS view of an HDF5 data file with the HDF5
  view.
  (a) Listing of a HDSv5 file using the standard HDS tracing tool,
  \textsc{hdstrace} \citep{SUN102}.
  (b) Listing of the HDF5 file corresponding to the HDS structures.
  The definition of the
  \texttt{H5T\_STRING} datatype has been elided for clarity. The
  listing was made using the standard \texttt{h5dump} command.
}
\label{fig:compare}
\end{figure}

\subsection{Primary locators}

In HDF5 the file is kept open until all identifiers associated with a
file are closed. HDS distinguishes primary identifiers from secondary
identifiers such that a file is closed when the count of active
primary locators reaches zero, even if some active secondary locators
remain. To implement this in
\new\ it is necessary to store every locator that is allocated in a
global data structure. We use the \texttt{uthash} macros \citep{uthash} to
implement a hash table indexed by the \texttt{hid\_t} HDF5 file identifier. Each
file identifier key then maps to a \texttt{utarray} dynamic array
containing the locators. The individual locators have a flag indicating
whether they are primary or secondary.
The \texttt{uthash} macros were
chosen since they did not require an additional library, they used a
BSD license that was compatible with the \new\ library license, and
the programming interface was reasonably straightforward.

When each locator is annulled these data structures are scanned to
check whether this locator was the final primary locator. If it is,
all remaining locators are themselves annulled. One complication is
that the file identifier returned will be different for each call to
\texttt{H5Fopen} so it is important to determine which file
identifiers are associated with the same file. Without accounting for
this, critical locators may be annulled at the wrong time. Rather than
attempt to guess what the HDF5 library has chosen to do by normalising
supplied filenames, the virtual file driver layer is queried to obtain
the Unix file descriptor. All file identifiers with a shared file
descriptor are queried before deciding whether a file should be
closed.

One consequence of this behind-the-scenes freeing of resources is that
it is possible for a library user that does not understand the
distinction between primary and secondary locators to be left with
pointers to structures that have been freed. To prevent unfortunate
crashes, when a locator is freed automatically the contents of the
structure are reset but the structure itself is not freed. This does
result in a small memory leak but is thought to be more acceptable
than a core dump.

\subsection{Locator groups}

Locator groups are not a feature of HDF5 and were implemented natively
in the \new\ library. The implementation is similar to the primary
locator system described previously except that the key for the
\texttt{uthash} mapping table is the group name rather than the file
identifier. When a group is flushed, all locators in the group are
annulled and the group is deleted from the hash table.

\subsection{Array slicing}

A very powerful feature of HDF5 is the concept of a dataspace. A
dataspace determines the rank and dimensions of a dataset and is used
to specify the size of the HDS primitives. When a slice or cell request is
made a single hyperslab selection is made which adjusts the external
view of the dataset. HDS slices and cells are much simpler than what
is possible in a hyperslab selection, and are restricted to simple
subsets of a region.

When a locator is vectorized the dataspace associated with the locator
is reshaped to be 1-dimensional. Subsequent slices of that vectorized
dataspace are then handled in the same way as before using a hyperslab.

\subsection{Type conversion}

HDF5 supports an extremely broad range of data types and automatic
conversion of numerical types when storing or retrieving a
dataset. Critically, HDF5 does not support type conversion of string
and logical types to numeric types (and vice versa) so this facility
has been added explicitly in the \new\ library to maintain
compatibility. This is simplified by HDS having the concept of a
``bad'' or ``magic'' value for each datatype that can be used to
indicate where a conversion was not possible.

\subsection{Memory mapping}

An important requirement for any HDS implementation is to
support direct memory mapping of files for both read and write
operations. This has worked well over the years and helps minimize
resource requirements. HDF5 has other priorities and advocates chunked
access to minimize resources rather than providing direct access to
the bytes on disk. The ability to split a dataset into multiple chunks
and to insert arbitrary compression filters and virtual file drivers
between the bytes trumps any perceived advantages of memory
mapping. In the \new\ implementation memory mapping is only attempted
if files are in read only mode, if HDF5 will return the byte offset to
the start of the dataset, if the HDF5 type system indicates that the
in-memory data type and the on-disk data types are compatible, and if
the virtual file driver will provide a file descriptor. In all other
cases memory is allocated using standard system calls when a user
requests memory mapping, and the data are written back to the file
when the data are ``unmapped''. Mapping can also be disabled using a
tuning parameter.

As a test a 4\,GB dataset was loaded into the GAIA visualization tool
\citep[][\ascl{1403.024}]{2009ASPC..411..575D}. With memory mapping enabled
the image displayed within two seconds and the process only used a
fews tens of megabytes of memory. With memory mapping disabled it took
about ten seconds to load the image and the process took 5\,GB of
memory.

The ability to memory map at all requires that datasets are created in
single chunks and are not resizable. This causes some problems with
HDS which assumes that all primitive objects are resizable. The \new\
dataset resizing function therefore attempts to use the native
HDF5 resizing function but is usually forced to create a new dataset
and copy the contents from the existing dataset, before deleting the
original and renaming the new dataset. This can result in significant
unused space in an HDF5 file.

\section{Implementation Issues}

The prototype library has largely shown that replacing native HDS with
an HDF5 implementation is feasible. Unfortunately we have found that
there are some incompatibilities that have required minor code changes
to Starlink applications. So far these have been restricted to
applications that open an input file with read access and then open an
output file with read/write access. If the input file and output file
are the same file (for example when copying a structure within a
file), HDS had no issue with this but in HDF5 this is strictly
forbidden due to the internal tracking of open files. The changes to
\textsc{Figaro} \citep[][\ascl{1411.022}]{1993ASPC...52..219S} and
\textsc{Hdstools} \citep{SUN245} result in the input being opened to
validate it but recording the full path to the requested object. Then
the input is closed and the output re-opened in read/write mode. Once
this happens the input file can be re-opened and the application can
continue as before. The modification also works with \old\ so can be
adopted at the expense of some more convoluted code.

When designing the mapping of HDS to HDF5 some care was taken to not
deliberately restrict the ability of the \new\ library to read HDF5
files that were not created by the library. To that end, attributes
were chosen that were either already in common usage, e.g.,
\texttt{CLASS}, or were chosen such that the absence of the attribute
would result in reasonable behavior (root naming and structure
dimensions). However, the implementation can not work miracles in
dealing with the mismatch between the HDS and HDF5 data models.
In particular, the HDS data model has no concept of attributes
in the sense that HDF5 has them. Figure~\ref{fig:h5traces} shows the
output of an HDS tracing program on a file created from a FITS file
as described by \citet{O4-4_adassxxiv}. HDS is able to read some of
the file contents but fails to read groups with names that exceed 16
characters. This limit can be increased by recompiling all Starlink
applications but HDS relies on this limit being fixed at compile time.
A more complex solution would be for HDS to return a shortened form of
the name to the HDS API, possibly keeping track of the mapping from
long name to short name internally. It is currently unclear how
important it will be to handle this situation. What is not obvious
from this trace is that the \texttt{HEADER} structure is not empty;
all the FITS headers are stored as attributes.

\begin{figure}
\small
\begin{verbatim}
HDF5ROOT  <HDFITS>

 PRIMARY      <HDU>      {structure}
   DATA(35,35)  <_WORD>  5419,5419,5332,6025,
                         ... 6659,6659,6572
   HEADER       <HDF5NATIVEGROUP> {structure}
      {structure is empty}

!! Invalid name string 'Photometric CALTABLE'
!  specified; more than 16 characters long
\end{verbatim}
\caption{Output from the HDS structure tracing application,
  \textsc{hdstrace} on an HDF5 file created without using the HDS
  library. HDS has read the \texttt{CLASS} attributes when available
  and replaced optional attribute values with placeholders for the
  name of the root group and the type of the \texttt{HEADER}
  structure. Unfortunately a group in the file has a name that exceeds
  the HDS fixed-width limit preventing HDS from accessing it. Furthermore,
  the \texttt{HEADER} structure is not in fact empty as all the FITS headers
  are actually stored as HDF5 attributes and these are invisible to the HDS
  data model.}
\label{fig:h5traces}
\end{figure}

\section{Metrics}

When considering adoption of a new format it is important to
consider any performance differences and whether the files use up
differing amounts of storage. These tests used HDF5 version 1.8.13 and
a late 2014 version of HDSv4.

\subsection{File Sizes}

\begin{table*}[ht]
  \caption{A comparison of file sizes resulting from identical
    operations where solely the file format is changed. Sizes are
    given for the natives files and gzipped versions. Also included
    are the sizes from using HDF5 native SHUFFLE/GZIP compression.
    The first two rows are from files that are continuously updated
    during processing; the remaining rows are from statically created
    data sets. All files sizes are in bytes.}
\label{tab:size}
\begin{center}
\begin{tabular}{lrrcrrcr}
\hline
File type &  \old\ & \new\  & v5/v4 & \old\ (gz) & \new (gz) & v5(gz)/v4(gz)
& HDF5 comp.\\ \hline

AGI database & 35\,328 & 191\,664 & 5.43 & 4\,473 & 9\,689& 2.17  & 403\,608\\
Parameter file & 5\,632 & 18\,600 & 3.30 & 745 & 1\,318 & 1.77 &  19\,800\\
SCUBA-2 & 18\,712\,576 & 18\,796\,530 & 1.00 & 15\,530\,536 &
15\,537\,267 & 1.00 & 14\,660\,769 \\
\KAPPA\ logo & 406\,016 & 411\,272 & 1.01 & 30\,633 & 31\,123 & 1.02 & 53\,827\\
\hline
\end{tabular}
\end{center}
\end{table*}

Test datasets were generated comparing the new format file sizes with
the original file sizes.\footnote{All these tests were done using the
  default file access property list settings. Selecting the latest
  format, via \texttt{H5Pset\_libver\_bounds}, results in slightly
  smaller files for three of the four tests but a larger file in the
  parameter file test. It has not yet been decided whether \new\ should
  adopt maximal backwards compatibility for files or always be on the cutting
  edge.}  A comparison is shown in
Table~\ref{tab:size}. The files were generated as follows:

\begin{enumerate}
\item The AGI graphics database generated by the \textsc{SpecDRE}
  demonstration script \citep[][\ascl{1407.003}]{1992StarB..10....8M}. The graphics database makes
  extensive use of arrays of structures and resizing of elements. The
  HDF5 variant is more than five times larger than the \old\ variant
  with 20\,\% of that accounted for by empty space.

\item An ADAM parameter file generated from the execution of the
  \textsc{ccdbig} \citep{SSN69} exercise script. 34\,\% of the HDF5
  file is empty space. Like the graphics database file, this file is
  updated constantly during program execution.

\item A SCUBA-2 acquisition file \citep{2014SPIE.9153E..03B}, which contains
  lots of data as well as table structures and is written in a single operation.

\item \KAPPA\ \citep[][\ascl{1403.022}]{SUN95} logo image consisting of a simple NDF
  with WCS and FITS header.
\end{enumerate}

The numbers indicate that for small files with many structures the
\new\ files are significantly larger. Some of this may be due to the
inability to resize datasets without deleting them but even if the
files are repacked they are still larger than \old\ versions.  For
larger data files the situation is less clear cut with the scientific
data dominating the file contents the overhead from HDF5 is much
lower. The advantages of HDF5 become obvious once native data
compression is used with the SCUBA-2 example file becoming 6\,\%
smaller than even the gzipped version of the \old\ format.

\subsection{Benchmarks}

The library has been tested on a number of standard Starlink benchmark
routines from \textsc{ccdpack} \citep[][\ascl{1403.021}]{SUN139},
\textsc{ccdbig}, and \textsc{starbench} \citep{SSN23}. An example data reduction
test was also executed using SCUBA-2 data and the \textsc{orac-dr}
pipeline \citep[][\ascl{1310.001}]{2015A&C.....9...40J}\footnote{These were
  observations 28, 31, 35, 44 and 51  from 2012 June 11th, reduced
  using the JCMT Science Archive public processing recipe
  \citep{2014SPIE9152-93}}. The results are shown in
Table~\ref{tab:bench} and indicate that for tasks using lots of small
files with lots of I/O \old\ is much faster. As the tests begin to use
more real-world processing tasks with larger datasets the difference
disappears and both libraries perform to within a few per cent. The
final test involving \textsc{orac-dr} indicates that there may be a
small performance advantage to not using memory mapping and this
result may inform later decisions on whether to switch to using
resizable chunked datasets in the future.

\begin{table}
\caption{Various Starlink benchmarking tests. All times are in
  seconds. The items with an asterisk indicate the benchmark was run
  with memory mapping disabled.}
\label{tab:bench}
\begin{center}
\begin{tabular}{lrrr}
\hline
   & \old\ & \new\ & v5/v4 \\ \hline
\textsc{starbench/specdre} & $0.82\pm0.02$ & $1.25\pm0.01$ & 1.52\\
\textsc{starbench/kappa} & $8.58\pm0.46$ & $9.88\pm0.09$ & 1.15\\
\textsc{ccdpack} & $5.02\pm0.08$ & $5.79\pm0.02$ & 1.15\\
\textsc{ccdbig} & $55.42\pm0.59$ & $54.04\pm0.07$ & 0.98\\
\textsc{ccdbig}(*) & $56.08\pm0.28$ & $54.64\pm0.19$ & 0.97\\
\textsc{orac-dr} & $486\pm5$ & $482\pm7$ & 0.99\\
\textsc{orac-dr}(*) & $450\pm2$ & $465\pm3$ & 1.03\\
\hline
\end{tabular}
\end{center}
\end{table}

\section{Conclusion}

A new HDF5-based implementation of the HDS programming interface has
been written which allows the Starlink software collection to be moved
to a more widely-used file format. All but a handful of the
approximately 150 HDS API functions have been implemented, with the
remaining few being the routines that query low level implementation
details. The \new\ library consists of approximately 10\,000 lines of
C with another 5\,000 lines of C for the implementation of the wrapper
(many of those lines are generated automatically). For comparison, the
\old\ library consisted of about 18\,000 lines of C and HDF5 itself
consists of about 120\,000 lines of C.

It has been shown that the library performs as well as the \old\
implementation in most tests involving reasonably-sized datasets and
opens up the possibility for Starlink data products to be more easily
consumed by others without requiring a format conversion. A native
Python interface to the HDS library does exist but it is far easier to
convince prospective consumers of the data files to use something
such as \texttt{h5py} \citep[e.g.,][]{pyhdf5} to read the data, albeit
with a different view of the data models. Furthermore, these files
would be readable by general HDF5 visualization tools.

The Starlink open-source community must now decide
whether to pursue this work and integrate it into the Starlink
software distribution. It is
possible that the project will decide to stick with \old\ and attempt
to update the library to support 64-bit dimension sizes. This is a
reasonable course of action to take, with an uncertain effort
requirement, although it does not solve the issues relating to lack of
documentation and sociological barrier to adoption of the Starlink
software. Furthermore, if the new implementation is adopted, serious
consideration must be made as to whether the approximately 4 million
HDS files in the JCMT Science Archive \citep{2015Economou} should be
converted to HDF5. There is a risk involved for the archive in terms
of the cost of keeping the old versions around and whether the
conversion has been done correctly. The benefit will be that the
raw data archive will immediately become more accessible to the
general astronomer.

This work also provides an alternative approach to porting FITS files to
HDF5 format \citep[see e.g.,][for other options]{O4-4_adassxxiv}. The Starlink
\textsc{convert} package \citep{1997STARB..19...14C,SUN55} has
received significant development effort over the years to map FITS to
a hierarchical data model. It will be interesting to see whether the
community can agree on a standard model for FITS to HDF5 conversion.

Now that a functioning prototype exists and has been proven to work
acceptably, we must consider the possibility of expanding the HDS API
to take advantage of HDF5 features. In particular compound datatypes
provide the prospect of native table access (the import of FITS binary tables would
benefit significantly from this), an updated slicing API could provide
access to hyperslabs, and the ability to specify chunking size and the
maximum expected size of a dataset could result in significant
efficiency benefits, albeit at the expense of memory mapping. It may
be possible to consider allowing the HDS and HDF5 APIs to be used
simultaneously on a single file. This has many attractions and
provides a simple path to enhancing native applications. It also would
mean it would be impossible to switch HDS from HDF5 to another format
in the future. If the Advanced Scientific Data Format
\citep[ASDF;][]{asdf,2015Greenfield} were to suddenly become popular in astronomy it would be
conceivable to investigate a port of the HDS API to ASDF. If HDF5
identifiers had been used natively in the code this would be a
significantly more complicated task. This is somewhat similar to the
problems that are faced in porting NDF to other formats. The NDF
standard \citep{SGP38} was specifically designed with an ``airlock''
API that allowed the user to obtain an HDS locator to extensions. This
flexibility was important in early adoption and provided an easy way
for extensions to be implemented. It also meant that any attempt to
switch NDF absolutely required that the HDS API was itself ported,
otherwise all the extensions in use would be unreadable. Indeed one key
motivation for this work is that it brings NDF along to HDF5 without
any NDF code or applications that use extensions having to be
modified.

\section{Acknowledgments}

This research has made use of NASA's Astrophysics Data System.  The
Starlink software is an open-source project that is maintained by the
Joint Astronomy Centre, Hawaii \citep{2014ASPC..485..391C}. I thank Peter Draper and David Berry
for testing development releases of the new library. I thank David
Berry, Danny Price, Rodney Warren-Smith and Mark Taylor for comments on the draft manuscript.
I thank the two referees for their comments on the paper and also
their comments on how to improve the library.

The source code for the new HDS library and the Starlink software
(\ascl{1110.012}) is open-source and is available on Github at
\htmladdnormallink{https://github.com/Starlink}.


\begin{thebibliography}{48}
\expandafter\ifx\csname natexlab\endcsname\relax\def\natexlab#1{#1}\fi
\providecommand{\url}[1]{\texttt{#1}}
\providecommand{\href}[2]{#2}
\providecommand{\path}[1]{#1}
\providecommand{\DOIprefix}{doi:}
\providecommand{\ArXivprefix}{arXiv:}
\providecommand{\URLprefix}{URL: }
\providecommand{\Pubmedprefix}{pmid:}
\providecommand{\doi}[1]{\href{http://dx.doi.org/#1}{\path{#1}}}
\providecommand{\Pubmed}[1]{\href{pmid:#1}{\path{#1}}}
\providecommand{\bibinfo}[2]{#2}
\ifx\xfnm\relax \def\xfnm[#1]{\unskip,\space#1}\fi
\bibitem[{{Alexov}  et~al.(2012){Alexov} et~al.}]{2012ASPC..461..283A}
\bibinfo{author}{{Alexov}, A.}, et~al., \bibinfo{year}{2012}.
\newblock \bibinfo{title}{{Status of LOFAR Data in HDF5 Format}}, in:
  \bibinfo{editor}{{Ballester}, P.}, \bibinfo{editor}{{Egret}, D.},
  \bibinfo{editor}{{Lorente}, N.P.F.} (Eds.), \bibinfo{booktitle}{Astronomical
  Data Analysis Software and Systems XXI}, volume \bibinfo{volume}{461} of
  \textit{\bibinfo{series}{\aspconf}}. p. \bibinfo{pages}{283}.
\bibitem[{{Allan}(1992)}]{1992ASPC...25..126A}
\bibinfo{author}{{Allan}, P.M.}, \bibinfo{year}{1992}.
\newblock \bibinfo{title}{{The ADAM software environment}}, in:
  \bibinfo{editor}{{Worrall}, D.M.}, \bibinfo{editor}{{Biemesderfer}, C.},
  \bibinfo{editor}{{Barnes}, J.} (Eds.), \bibinfo{booktitle}{Astronomical Data
  Analysis Software and Systems I}, volume~\bibinfo{volume}{25} of
  \textit{\bibinfo{series}{\aspconf}}. p. \bibinfo{pages}{126}.
\bibitem[{Beard et~al.(2006)Beard, Allan, Currie and Draper}]{SUN7}
\bibinfo{author}{Beard, S.M.}, \bibinfo{author}{Allan, P.M.},
  \bibinfo{author}{Currie, M.J.}, \bibinfo{author}{Draper, P.W.},
  \bibinfo{year}{2006}.
\newblock \bibinfo{title}{{GENERIC --- A utility for preprocessing generic
  Fortran and C subroutines}}.
\newblock \bibinfo{type}{Starlink User Note} \bibinfo{number}{7}. Starlink
  Project.
\bibitem[{{Bell} et~al.(2014){Bell}, {Currie}, {Redman}, {Purves} and
  {Jenness}}]{2014ASPC..485..143B}
\bibinfo{author}{{Bell}, G.S.}, \bibinfo{author}{{Currie}, M.J.},
  \bibinfo{author}{{Redman}, R.O.}, \bibinfo{author}{{Purves}, M.},
  \bibinfo{author}{{Jenness}, T.}, \bibinfo{year}{2014}.
\newblock \bibinfo{title}{{A New Archive of UKIRT Legacy Data at CADC}}, in:
  \bibinfo{editor}{{Manset}, N.}, \bibinfo{editor}{{Forshay}, P.} (Eds.),
  \bibinfo{booktitle}{Astronomical Data Analysis Software and Systems XXIII},
  volume \bibinfo{volume}{485} of \textit{\bibinfo{series}{ASP Conf.\ Ser.}}.
  p. \bibinfo{pages}{143}.
\bibitem[{Bell et~al.(2014)Bell, Graves, Currie, Berry, Parsons, Jenness,
  Redman, Dempsey, Johnstone and Economou}]{2014SPIE9152-93}
Bell, G.~S. et~al., \bibinfo{year}{2014}.
\newblock \bibinfo{title}{{Advanced data products for the JCMT Science
  Archive}}, in: \bibinfo{editor}{Chiozzi, G.}, \bibinfo{editor}{Radziwill,
  N.M.} (Eds.), \bibinfo{booktitle}{{Software and Cyberinfrastructure for
  Astronomy III}}, volume \bibinfo{volume}{9152} of
  \textit{\bibinfo{series}{\procspie}}. p. \bibinfo{pages}{91522J}.
\newblock \DOIprefix\doi{10.1117/12.2054983}.
\bibitem[{{Bintley} et~al.(2014){Bintley}, {Holland}, {MacIntosh}, {Friberg},
  {Bell}, {Berke}, {Berry}, {Berthold}, {Cookson}, {Coulson}, {Currie},
  {Dempsey}, {Gibb}, {Gorges}, {Graves}, {Jenness}, {Johnstone}, {Parsons},
  {Thomas}, {Walther} and {Wouterloot}}]{2014SPIE.9153E..03B}
{Bintley}, D. et~al., \bibinfo{year}{2014}.
\newblock \bibinfo{title}{{SCUBA-2: an update on the performance of the 10,000
  pixel bolometer camera after two years of science operation at the JCMT}},
  in: \bibinfo{editor}{Holland, W.S.}, \bibinfo{editor}{Zmuidzinas, J.} (Eds.),
  \bibinfo{booktitle}{Millimeter, Submillimeter, and Far-Infrared Detectors and
  Instrumentation for Astronomy VII}, volume \bibinfo{volume}{9153} of
  \textit{\bibinfo{series}{\procspie}}. p. \bibinfo{pages}{915303}.
\newblock \DOIprefix\doi{10.1117/12.2055231}.
\bibitem[{de~Buyl et~al.(2014)de~Buyl, Colberg and Höfling}]{deBuyl20141546}
\bibinfo{author}{de~Buyl, P.}, \bibinfo{author}{Colberg, P.H.},
  \bibinfo{author}{Höfling, F.}, \bibinfo{year}{2014}.
\newblock \bibinfo{title}{{H5MD: A structured, efficient, and portable file
  format for molecular data}}.
\newblock \bibinfo{journal}{Computer Physics Communications}
  \bibinfo{volume}{185}, \bibinfo{pages}{1546 -- 1553}.
\newblock \DOIprefix\doi{10.1016/j.cpc.2014.01.018}.
\bibitem[{Chipperfield(2002)}]{SUN245}
\bibinfo{author}{Chipperfield, A.J.}, \bibinfo{year}{2002}.
\newblock \bibinfo{title}{{HDSTOOLS --- Tools to Display and Edit HDS
  Objects}}.
\newblock \bibinfo{type}{Starlink User Note} \bibinfo{number}{245}. Starlink
  Project.
\bibitem[{Collette(2013)}]{pyhdf5}
\bibinfo{author}{Collette, A.}, \bibinfo{year}{2013}.
\newblock \bibinfo{title}{Python and HDF5}.
\newblock \bibinfo{publisher}{O'Reilly Media, Inc.}
\newblock \bibinfo{note}{ISBN: 978-1-449-36783-1}.
\bibitem[{Currie(1997)}]{1997STARB..19...14C}
\bibinfo{author}{Currie, M.J.}, \bibinfo{year}{1997}.
\newblock \bibinfo{title}{Data-format conversion}.
\newblock \bibinfo{journal}{Starlink Bulletin} \bibinfo{volume}{19},
  \bibinfo{pages}{14}.
\bibitem[{Currie(2014)}]{SUN102}
\bibinfo{author}{Currie, M.J.}, \bibinfo{year}{2014}.
\newblock \bibinfo{title}{{HDSTRACE --- Listing HDS Data Files}}.
\newblock \bibinfo{type}{Starlink User Note} \bibinfo{number}{102}. Starlink
  Project.
\bibitem[{Currie and Berry(2013)}]{SUN95}
\bibinfo{author}{Currie, M.J.}, \bibinfo{author}{Berry, D.S.},
  \bibinfo{year}{2013}.
\newblock \bibinfo{title}{{KAPPA -- Kernel Application Package}}.
\newblock \bibinfo{type}{Starlink User Note} \bibinfo{number}{95}. Starlink
  Project.
\bibitem[{{Currie} et~al.(2014){Currie}, {Berry}, {Jenness}, {Gibb}, {Bell} and
  {Draper}}]{2014ASPC..485..391C}
\bibinfo{author}{{Currie}, M.J.}, \bibinfo{author}{{Berry}, D.S.},
  \bibinfo{author}{{Jenness}, T.}, \bibinfo{author}{{Gibb}, A.G.},
  \bibinfo{author}{{Bell}, G.S.}, \bibinfo{author}{{Draper}, P.W.},
  \bibinfo{year}{2014}.
\newblock \bibinfo{title}{{Starlink Software in 2013}}, in:
  \bibinfo{editor}{{Manset}, N.}, \bibinfo{editor}{{Forshay}, P.} (Eds.),
  \bibinfo{booktitle}{Astronomical Data Analysis Software and Systems XXIII},
  volume \bibinfo{volume}{485} of \textit{\bibinfo{series}{ASP Conf.\ Ser.}}.
  p. \bibinfo{pages}{391}.
\bibitem[{Currie et~al.(1996)Currie, Privett, Chipperfield, Berry and
  Davenhall}]{SUN55}
\bibinfo{author}{Currie, M.J.}, \bibinfo{author}{Privett, G.J.},
  \bibinfo{author}{Chipperfield, A.J.}, \bibinfo{author}{Berry, D.S.},
  \bibinfo{author}{Davenhall, A.C.}, \bibinfo{year}{1996}.
\newblock \bibinfo{title}{{CONVERT -- A Format-conversion Package}}.
\newblock \bibinfo{type}{Starlink User Note} \bibinfo{number}{55}. Starlink
  Project.
\bibitem[{Currie et~al.(1988)Currie, Wallace and Warren-Smith}]{SGP38}
\bibinfo{author}{Currie, M.J.}, \bibinfo{author}{Wallace, P.T.},
  \bibinfo{author}{Warren-Smith, R.F.}, \bibinfo{year}{1988}.
\newblock \bibinfo{title}{Starlink Standard Data Structures}.
\newblock \bibinfo{type}{Starlink General Paper} \bibinfo{number}{38}. Starlink
  Project.
\bibitem[{{Disney} and {Wallace}(1982)}]{1982QJRAS..23..485D}
\bibinfo{author}{{Disney}, M.J.}, \bibinfo{author}{{Wallace}, P.T.},
  \bibinfo{year}{1982}.
\newblock \bibinfo{title}{{STARLINK}}.
\newblock \bibinfo{journal}{\qjras} \bibinfo{volume}{23}, \bibinfo{pages}{485}.
\bibitem[{Dougherty et~al.(2009)Dougherty, Folk, Zadok, Bernstein, Bernstein,
  Eliceiri, Benger and Best}]{Dougherty:2009:UBI:1562764.1562781}
\bibinfo{author}{Dougherty, M.T.}, \bibinfo{author}{Folk, M.J.},
  \bibinfo{author}{Zadok, E.}, \bibinfo{author}{Bernstein, H.J.},
  \bibinfo{author}{Bernstein, F.C.}, \bibinfo{author}{Eliceiri, K.W.},
  \bibinfo{author}{Benger, W.}, \bibinfo{author}{Best, C.},
  \bibinfo{year}{2009}.
\newblock \bibinfo{title}{{Unifying Biological Image Formats with HDF5}}.
\newblock \bibinfo{journal}{Commun. ACM} \bibinfo{volume}{52},
  \bibinfo{pages}{42--47}.
\newblock \DOIprefix\doi{10.1145/1562764.1562781}.
\bibitem[{{Draper} et~al.(2009){Draper}, {Berry}, {Jenness} and
  {Economou}}]{2009ASPC..411..575D}
\bibinfo{author}{{Draper}, P.W.}, \bibinfo{author}{{Berry}, D.S.},
  \bibinfo{author}{{Jenness}, T.}, \bibinfo{author}{{Economou}, F.},
  \bibinfo{year}{2009}.
\newblock \bibinfo{title}{{GAIA -- Version 4}}, in:
  \bibinfo{editor}{{Bohlender}, D.A.}, \bibinfo{editor}{{Durand}, D.},
  \bibinfo{editor}{{Dowler}, P.} (Eds.), \bibinfo{booktitle}{Astronomical Data
  Analysis Software and Systems XVIII}, volume \bibinfo{volume}{411} of
  \textit{\bibinfo{series}{\aspconf}}. p. \bibinfo{pages}{575}.
\bibitem[{Draper et~al.(2011)Draper, Taylor and Allan}]{SUN139}
\bibinfo{author}{Draper, P.W.}, \bibinfo{author}{Taylor, M.B.},
  \bibinfo{author}{Allan, A.}, \bibinfo{year}{2011}.
\newblock \bibinfo{title}{{CCDPACK -- CCD data reduction package}}.
\newblock \bibinfo{type}{Starlink User Note} \bibinfo{number}{139}. Starlink
  Project.
\bibitem[{Droettboom and Bray(2014)}]{asdf}
\bibinfo{author}{Droettboom, M.}, \bibinfo{author}{Bray, E.},
  \bibinfo{year}{2014}.
\newblock \bibinfo{title}{ASDF Standard}.
\newblock \bibinfo{type}{Technical Report}. Space Telescope Science Institute.
\newblock \URLprefix \url{http://asdf-standard.readthedocs.org/en/latest/}.
\bibitem[{Eaton and McIlwrath(2014)}]{SUN48}
\bibinfo{author}{Eaton, N.}, \bibinfo{author}{McIlwrath, B.},
  \bibinfo{year}{2014}.
\newblock \bibinfo{title}{{AGI -- Applications Graphics Interface Library}}.
\newblock \bibinfo{type}{Starlink User Note} \bibinfo{number}{48}. Starlink
  Project.
\bibitem[{Economou et~al.(2015)Economou, Gaudet, Jenness, Redman, Goliath,
  Dowler, Currie, Bell, Graves, Ouellette, Schade and
  Chrysostomou}]{2015Economou}
Economou, F. et~al., \bibinfo{year}{2015}.
\newblock \bibinfo{title}{{Observatory/data centre partnerships and the
  VO-centric archive: The JCMT Science Archive experience}}.
\newblock \bibinfo{journal}{Astron.\ Comp.} \bibinfo{volume}{in press}.
\newblock \DOIprefix\doi{10.1016/j.ascom.2014.12.005},
  \href{http://arxiv.org/abs/1412.4399}{{\tt arXiv:1412.4399}}.
\bibitem[{Folk(2010)}]{Folk2010}
\bibinfo{author}{Folk, M.}, \bibinfo{year}{2010}.
\newblock \bibinfo{title}{{HDF --- Past, Present, Future}}.
\newblock \bibinfo{howpublished}{HDF and HDF-EOS Workshop XIV}.
\newblock \URLprefix
  \url{http://www.slideshare.net/HDFEOS/the-hdf-group-past-present-and-future}.
\bibitem[{Greenfield et~al.(2015)Greenfield, Droettboom and
  Bray}]{2015Greenfield}
\bibinfo{author}{Greenfield, P.}, \bibinfo{author}{Droettboom, M.},
  \bibinfo{author}{Bray, E.}, \bibinfo{year}{2015}.
\newblock \bibinfo{title}{{ASDF: a new data format for astronomy}}.
\newblock \bibinfo{journal}{Astron.\ Comp.} \bibinfo{volume}{in press}.
\bibitem[{{Greisen} et~al.(1980){Greisen}, {Wells} and
  {Harten}}]{1980SPIE..264..298G}
\bibinfo{author}{{Greisen}, E.W.}, \bibinfo{author}{{Wells}, D.C.},
  \bibinfo{author}{{Harten}, R.H.}, \bibinfo{year}{1980}.
\newblock \bibinfo{title}{{The FITS Tape Formats: Flexible Image Transport
  Systems}}, in: \bibinfo{editor}{{Elliott}, D.A.} (Ed.),
  \bibinfo{booktitle}{Applications of Digital Image Processing to Astronomy},
  volume \bibinfo{volume}{264} of \textit{\bibinfo{series}{Proc.\ SPIE}}. p.
  \bibinfo{pages}{298}.
\newblock \DOIprefix\doi{10.1117/12.959819}.
\bibitem[{Hanson(2014)}]{uthash}
\bibinfo{author}{Hanson, T.D.}, \bibinfo{year}{2014}.
\newblock \bibinfo{title}{\texttt{uthash} User Guide}.
\newblock \URLprefix \url{https://troydhanson.github.io/uthash/userguide.html}.
\bibitem[{{Harten} et~al.(1988){Harten}, {Grosb{\o}l}, {Greisen} and
  {Wells}}]{1988A&AS...73..365H}
\bibinfo{author}{{Harten}, R.H.}, \bibinfo{author}{{Grosb{\o}l}, P.},
  \bibinfo{author}{{Greisen}, E.W.}, \bibinfo{author}{{Wells}, D.C.},
  \bibinfo{year}{1988}.
\newblock \bibinfo{title}{{The FITS tables extension}}.
\newblock \bibinfo{journal}{\aaps} \bibinfo{volume}{73},
  \bibinfo{pages}{365--372}.
\bibitem[{Jenness  et~al.(2015)Jenness et~al.}]{2015Jenness}
\bibinfo{author}{Jenness, T.}, et~al., \bibinfo{year}{2015}.
\newblock \bibinfo{title}{{Learning from 25 years of the extensible
  \emph{N}-Dimensional Data Format}}.
\newblock \bibinfo{journal}{Astron.\ Comp.} \bibinfo{volume}{in press}.
\newblock \DOIprefix\doi{10.1016/j.ascom.2014.11.001},
  \href{http://arxiv.org/abs/1410.7513}{{\tt arXiv:1410.7513}}.
\bibitem[{{Jenness} and {Economou}(2011)}]{2011tfa..confE..42J}
\bibinfo{author}{{Jenness}, T.}, \bibinfo{author}{{Economou}, F.},
  \bibinfo{year}{2011}.
\newblock \bibinfo{title}{{Data Management at the UKIRT and JCMT}}, in:
  {Gajadhar}, S. et~al. (Eds.), \bibinfo{booktitle}{Telescopes from Afar},
  p.~\bibinfo{pages}{42}.
\newblock \href{http://arxiv.org/abs/1111.5855}{{\tt arXiv:1111.5855}}.
\bibitem[{Jenness and Economou(2015)}]{2015A&C.....9...40J}
\bibinfo{author}{Jenness, T.}, \bibinfo{author}{Economou, F.},
  \bibinfo{year}{2015}.
\newblock \bibinfo{title}{{ORAC-DR: A generic data reduction pipeline
  infrastructure}}.
\newblock \bibinfo{journal}{Astron.\ Comp} \bibinfo{volume}{9},
  \bibinfo{pages}{40}.
\newblock \DOIprefix\doi{10.1016/j.ascom.2014.10.005},
  \href{http://arxiv.org/abs/1410.7509}{{\tt arXiv:1410.7509}}.
\bibitem[{Jenness et~al.(2014)Jenness, Shepherd, Schaaf, Sayers, Ossenkopf,
  Nikola, Marsden, Higgins, Edwards and Brazier}]{2014SPIE.9152E..2WJ}
Jenness, T. et~al., \bibinfo{year}{2014}.
\newblock \bibinfo{title}{{An overview of the planned CCAT software system}},
  in: \bibinfo{editor}{Chiozzi, G.}, \bibinfo{editor}{Radziwill, N.M.} (Eds.),
  \bibinfo{booktitle}{Software and Cyberinfrastructure for Astronomy III},
  volume \bibinfo{volume}{9152} of \textit{\bibinfo{series}{Proc.\ SPIE}}. p.
  \bibinfo{pages}{91522W}.
\newblock \DOIprefix\doi{10.1117/12.2056516},
  \href{http://arxiv.org/abs/1406.1515}{{\tt arXiv:1406.1515}}.
\bibitem[{Krauskopf and Paulsen(1988)}]{HDF1}
\bibinfo{author}{Krauskopf, T.}, \bibinfo{author}{Paulsen, G.B.},
  \bibinfo{year}{1988}.
\newblock \bibinfo{title}{Hierarchical Data Format Version 1.0}.
\newblock \bibinfo{type}{Technical Report}. National Center for Supercomputing
  Applications, University of Illinois at Urbana-Champaign.
\bibitem[{Lawden(1991)}]{1991STARB...8....2L}
\bibinfo{author}{Lawden, M.D.}, \bibinfo{year}{1991}.
\newblock \bibinfo{title}{Ten years ago}.
\newblock \bibinfo{journal}{Starlink Bulletin} \bibinfo{volume}{8},
  \bibinfo{pages}{2}.
\bibitem[{Lupton(1989)}]{SSN27a}
\bibinfo{author}{Lupton, W.F.}, \bibinfo{year}{1989}.
\newblock \bibinfo{title}{{HDS -- Hierarchical Data System}}.
\newblock \bibinfo{type}{Starlink System Note} \bibinfo{number}{27}. Starlink
  Project.
\bibitem[{{Meyerdierks}(1992)}]{1992StarB..10....8M}
\bibinfo{author}{{Meyerdierks}, H.}, \bibinfo{year}{1992}.
\newblock \bibinfo{title}{{Spectroscopy with SPECDRE}}.
\newblock \bibinfo{journal}{Starlink Bulletin} \bibinfo{volume}{10},
  \bibinfo{pages}{8--9}.
\bibitem[{Meyerdierks(1993)}]{1993STARB..12...10M}
\bibinfo{author}{Meyerdierks, H.}, \bibinfo{year}{1993}.
\newblock \bibinfo{title}{{Unix Figaro? --- Portable Figaro!}}
\newblock \bibinfo{journal}{Starlink Bulletin} \bibinfo{volume}{12},
  \bibinfo{pages}{10}.
\bibitem[{Mink et~al.(2015)Mink, Mann, Hanisch, Rots, Seaman, Jenness and
  Thomas}]{B1_adassxxiv}
\bibinfo{author}{Mink, J.}, \bibinfo{author}{Mann, R.G.},
  \bibinfo{author}{Hanisch, R.}, \bibinfo{author}{Rots, A.},
  \bibinfo{author}{Seaman, R.}, \bibinfo{author}{Jenness, T.},
  \bibinfo{author}{Thomas, B.}, \bibinfo{year}{2015}.
\newblock \bibinfo{title}{{The Past, Present and Future of Astronomical Data
  Formats}}, in: \bibinfo{editor}{Taylor, A.R.}, \bibinfo{editor}{Stil, J.M.}
  (Eds.), \bibinfo{booktitle}{Astronomical Data Analysis Software and Systems
  XXIV}, volume \bibinfo{volume}{in press} of \textit{\bibinfo{series}{ASP
  Conf.\ Ser.}}
\newblock \href{http://arxiv.org/abs/1411.0996}{{\tt arXiv:1411.0996}}.
\bibitem[{Pedersen et~al.(2013)Pedersen, Rees, Basham and
  Ferner}]{1742-6596-425-6-062008}
\bibinfo{author}{Pedersen, U.K.}, \bibinfo{author}{Rees, N.},
  \bibinfo{author}{Basham, M.}, \bibinfo{author}{Ferner, F.J.K.},
  \bibinfo{year}{2013}.
\newblock \bibinfo{title}{Handling high data rate detectors at diamond light
  source}.
\newblock \bibinfo{journal}{Journal of Physics: Conference Series}
  \bibinfo{volume}{425}, \bibinfo{pages}{062008}.
\newblock \DOIprefix\doi{10.1088/1742-6596/425/6/062008}.
\bibitem[{Price et~al.(2015)Price, Greenhill and Barsdell}]{O4-4_adassxxiv}
\bibinfo{author}{Price, D.}, \bibinfo{author}{Greenhill, L.},
  \bibinfo{author}{Barsdell, B.}, \bibinfo{year}{2015}.
\newblock \bibinfo{title}{{Is HDF5 a good format to replace UVFITS?}}, in:
  \bibinfo{editor}{Taylor, A.R.}, \bibinfo{editor}{Stil, J.M.} (Eds.),
  \bibinfo{booktitle}{Astronomical Data Analysis Software and Systems XXIV},
  volume \bibinfo{volume}{in press} of \textit{\bibinfo{series}{\aspconf}}.
\newblock \href{http://arxiv.org/abs/1410.8788}{{\tt arXiv:1410.8788}}.
\bibitem[{Rankin et~al.(2003)Rankin, Bly, Gledhill and Clayton}]{SSN23}
\bibinfo{author}{Rankin, S.E.}, \bibinfo{author}{Bly, M.J.},
  \bibinfo{author}{Gledhill, T.M..}, \bibinfo{author}{Clayton, C.A.},
  \bibinfo{year}{2003}.
\newblock \bibinfo{title}{{Starlink Benchmarking Utility}}.
\newblock \bibinfo{type}{Starlink System Note} \bibinfo{number}{23}. Starlink
  Project.
\bibitem[{Rees et~al.(2008)Rees, Chipperfield and Draper}]{SSN4}
\bibinfo{author}{Rees, P.C.T.}, \bibinfo{author}{Chipperfield, A.J.},
  \bibinfo{author}{Draper, P.W.}, \bibinfo{year}{2008}.
\newblock \bibinfo{title}{{EMS --- Error Message Service Programmer's Manual}}.
\newblock \bibinfo{type}{Starlink System Note} \bibinfo{number}{4}. Starlink
  Project.
\bibitem[{Rew and Hartnett(2004)}]{2004Rew}
\bibinfo{author}{Rew, R.K.}, \bibinfo{author}{Hartnett, E.J.},
  \bibinfo{year}{2004}.
\newblock \bibinfo{title}{{Merging NetCDF and HDF5}}, in:
  \bibinfo{booktitle}{21st International Conference on Interactive Information
  Processing Systems (IIPS) for Meteorology, Oceanography, and Hydrology}, p.
  \bibinfo{pages}{P1.11}.
\newblock \URLprefix \url{https://ams.confex.com/ams/pdfpapers/73771.pdf}.
\bibitem[{Schaaf et~al.(2015)Schaaf, Brazier, Jenness, Nikola and
  Shepherd}]{P3-1_adassxxiv}
\bibinfo{author}{Schaaf, R.}, \bibinfo{author}{Brazier, A.},
  \bibinfo{author}{Jenness, T.}, \bibinfo{author}{Nikola, T.},
  \bibinfo{author}{Shepherd, M.}, \bibinfo{year}{2015}.
\newblock \bibinfo{title}{{A new HDF5 based raw data model for CCAT}}, in:
  \bibinfo{editor}{Taylor, A.R.}, \bibinfo{editor}{Stil, J.M.} (Eds.),
  \bibinfo{booktitle}{Astronomical Data Analysis Software and Systems XXIV},
  volume \bibinfo{volume}{in press} of \textit{\bibinfo{series}{\aspconf}}.
\newblock \href{http://arxiv.org/abs/1410.8788}{{\tt arXiv:1410.8788}}.
\bibitem[{{Shortridge}(1993)}]{1993ASPC...52..219S}
\bibinfo{author}{{Shortridge}, K.}, \bibinfo{year}{1993}.
\newblock \bibinfo{title}{{The Evolution of the FIGARO Data Reduction System}},
  in: \bibinfo{editor}{{Hanisch}, R.J.}, \bibinfo{editor}{{Brissenden},
  R.J.V.}, \bibinfo{editor}{{Barnes}, J.} (Eds.),
  \bibinfo{booktitle}{Astronomical Data Analysis Software and Systems II},
  volume~\bibinfo{volume}{52} of \textit{\bibinfo{series}{ASP Conf.\ Ser.}}. p.
  \bibinfo{pages}{219}.
\bibitem[{Taylor(1998)}]{SSN69}
\bibinfo{author}{Taylor, M.B.}, \bibinfo{year}{1998}.
\newblock \bibinfo{title}{{CCDBIG: Assessing CCDPACK resource usage for large
  data sets}}.
\newblock \bibinfo{type}{Starlink System Note} \bibinfo{number}{69}. Starlink
  Project.
\bibitem[{Thomas  et~al.(2015)Thomas et~al.}]{2015Thomas}
\bibinfo{author}{Thomas, B.}, et~al., \bibinfo{year}{2015}.
\newblock \bibinfo{title}{{The Future of Astronomical Data Formats: Learning
  from FITS}}.
\newblock \bibinfo{journal}{Astron.\ Comp.} \bibinfo{volume}{in press}.
\newblock \DOIprefix\doi{10.1016/j.ascom.2015.01.009},
  \href{http://arxiv.org/abs/1502.00996}{{\tt arXiv:1502.00996}}.
\bibitem[{{Wells} et~al.(1981){Wells}, {Greisen} and
  {Harten}}]{1981A&AS...44..363W}
\bibinfo{author}{{Wells}, D.C.}, \bibinfo{author}{{Greisen}, E.W.},
  \bibinfo{author}{{Harten}, R.H.}, \bibinfo{year}{1981}.
\newblock \bibinfo{title}{{FITS -- a Flexible Image Transport System}}.
\newblock \bibinfo{journal}{\aaps} \bibinfo{volume}{44}, \bibinfo{pages}{363}.
\bibitem[{Yang et~al.(2005)Yang, McGrath and Folk}]{2005Yang}
\bibinfo{author}{Yang, M.}, \bibinfo{author}{McGrath, R.E.},
  \bibinfo{author}{Folk, M.}, \bibinfo{year}{2005}.
\newblock \bibinfo{title}{{HDF5 --- A High performance data Format for Earth
  Science}}, in: \bibinfo{booktitle}{21st International Conference on
  Interactive Information Processing Systems (IIPS) for Meteorology,
  Oceanography, and Hydrology}, p. \bibinfo{pages}{P2.42}.
\newblock \URLprefix \url{https://ams.confex.com/ams/pdfpapers/88475.pdf}.

\end{thebibliography}
\end{document}